\documentclass[%
 reprint,
 amsmath,amssymb,
 aps,
 pra,
]{revtex4-2}

\usepackage{graphicx}
\usepackage{dcolumn}
\usepackage{bm}
\usepackage{hyperref}

\newcommand{\dif}{\mathrm{d}}

\begin{document}

\preprint{APS/123-QED}

\title{Topological photonic Su-Schrieffer-Heeger-type coupler}

\author{Nikolaos K. Efremidis}
 \email{nefrem@uoc.gr}

\affiliation{Department of Mathematics and Applied Mathematics, University of Crete, 70013 Heraklion, Crete, Greece}
\affiliation{%
  Institute of Applied and Computational Mathematics, Foundation for Research and Technology-Hellas (FORTH), 70013 Heraklion, Crete, Greece
}%

\date{\today}

\begin{abstract}
  We examine the coupling process between the surface modes of a Su-Schrieffer-Heeger lattice both in the linear and the nonlinear regimes. We first develop a coupled-mode theory formalism for the modes of a finite lattice with zero boundary conditions. Our analysis relies on the closed-form expressions for the bulk and the surface eigenmodes of the system. The coupled-mode theory formalism is based on a decomposition of the supermodes into sublattice modes. In the case of the two zero sublattice surface modes, this leads to periodic oscillations between them without the involvement of the bulk modes. We analytically show that launching light only on the waveguide that is located at either edge of the array, can be very effective in successfully exciting the respective surface mode. We extend our analysis, in the case of Kerr nonlinearity, and develop a simplified model that accounts only for the surface modes. By direct numerical simulations, we find that this model can very accurately capture the dynamics of the surface modes when the nonlinearity is small or moderate. On the other hand, in the case of strong nonlinearity, wave mixing leads to the quasi-periodic excitation of the bulk modes, or even to a chaotic behavior where all the modes of the systems are excited, and no prominent signature of the surface modes can be detected. 
\end{abstract}

\maketitle

\section{Introduction}

Topological phenomena, such as the quantum Hall effect, play a fundamental role in condensed-matter physics~\cite{hasan-rmp2010,asbot-springer2016}. In optics, topological directional interface modes were first predicted in~\cite{halda-prl2008,raghu-pra2008}. In~\cite{wang-nature2009}, such topologically protected states were observed using magneto-optical photonic crystals. Even in the absence of a magnetic field, photonic Floquet topological insulators were experimentally observed, by breaking the time-reversal symmetry of the system~\cite{recht-nature2013}. Since then, a significant amount of work has been devoted in the field of topological photonics~\cite{lu-np2014,ozawa-rmp2019}, including topological states in ring resonators~\cite{hafez-np2013}, and topological insulator lasers~\cite{harar-science2018,bandr-science2018}.

In the case of one transverse dimension, the Su-Schrieffer-Heeger (SSH) lattice is known to support interface modes of topological origin~\cite{su-prl1979}. In particular, the SSH model is a diatomic lattice with alternating coupling coefficients. It has a topological winding number which is equal to 1, when the intercell coupling coefficient is stronger than the intracell coupling coefficient, and zero in the opposite case. Surface modes of the SSH lattice were experimentally observed in optically induced photonic lattices~\cite{malko-ol2009,malko-pra2009}. Topological interface states were also observed in Aubry-Andr\'e quasicrystals~\cite{kraus-prl2012}, as well as in photonic quantum walks~\cite{kitag-nc2012}. The presence of gain and loss in systems with alternating coupling coefficients has been utilized to observe topological transitions~\cite{zeune-prl2015}, midgap states~\cite{schom-ol2013}, optical isolation~\cite{ganai-ol2015}, enhancement of topologically induced interface states in microresonators~\cite{poli-nc2015}, and recovery of the topological zero modes in non-Hermitian lattices~\cite{song-prl2019}. In the presence of nonlinearity, the surface modes of the SSH lattice turn into a family of topological gap solitons~\cite{malko-pra2009,meier-nc2016,guo-ol2020}.

Most commonly, directional light coupling is achieved between optical fibers or waveguides in the total internal reflection regime~\cite{agrawal-applications}. Light coupling in non-topological systems is also possible by using different mechanisms, for example, Bragg reflection~\cite{yeh-oc1976,efrem-oe2005}, adiabatic elimination methods~\cite{mreje-nc2015,mreje-nl2015}, or even engineered lattices that support fractional revivals~\cite{efrem-oc2005}. In a topological setting, Floquet engineering has been examined in~\cite{pan-arxiv2020}, and adiabatic pumping of light in Aubry-Andr\'e lattices, between the two surface modes, has been observed in~\cite{kraus-prl2012}, in a process that relies on dynamically modifying the Hamiltonian, as a function of the propagation distance. Topological adiabatic pumping processes have also been examined in~\cite{verbi-prb2015,zilber-nature2018,cerja-light2020}.
In the SSH lattice, it is known that in the linear case periodic oscillations occur between the ``zero modes'', which are the local surface modes located on the left and the right side of the lattice (see for example~\cite{song-prl2019}). However, a theoretical framework that examines the coupling process and its efficiency is missing.

In this paper, we examine the coupling between the modes of an SSH lattice both in the linear and in the nonlinear regimes. We utilize the closed-form expressions for the modes of the SSH lattice, that were originally derived in~\cite{delpl-prb2011}. We develop a coupled-mode theory formalism that applies to both the bulk and the surface modes. By decomposing the supermodes into local sublattice modes, we find that coupling processes take place between pairs of modes with different sublattice symmetry. Of particular interest are the periodic oscillations between the surface zero modes, which are localized on the left and the right side of the lattice.
In experiments, the exact initial excitation of the surface modes might be a difficult task. In this respect, we analytically investigate the amount of power that excites the surface and the bulk sublattice modes, when only one waveguide on the left (or, equivalently, the right) edge of the array is excited. We find that the efficiency of the surface mode excitation in this very simple scheme can be very high. We extend our analysis in the nonlinear regime by accounting for the effect of Kerr nonlinearity and perform a series of numerical simulations. We also derive a simplified coupler-type model that accounts only for the surface zero modes. The comparison of this model, with the full SSH system is found to be excellent in the case of small and even moderate nonlinearities. By further increasing the nonlinearity, wave mixing can lead to the quasiperiodic excitation of the bulk modes. Finally, in the strongly nonlinear regime, there is a transition length, after which the behavior becomes chaotic with all the modes being excited, and no signature of the presence of the surface modes. The transition length decreases as the nonlinearity of the system increases. 

A main advantage in using such topological couplers as compared to conventional couplers, is related to the fact that the former is an array configuration consisting of a variable number of elements. Thus, we can select to couple light between distant non-adjacent lattice elements, a utilization that can be useful in multiport switching and routing applications.

\begin{figure}
  \centering
  \includegraphics[width=\columnwidth]{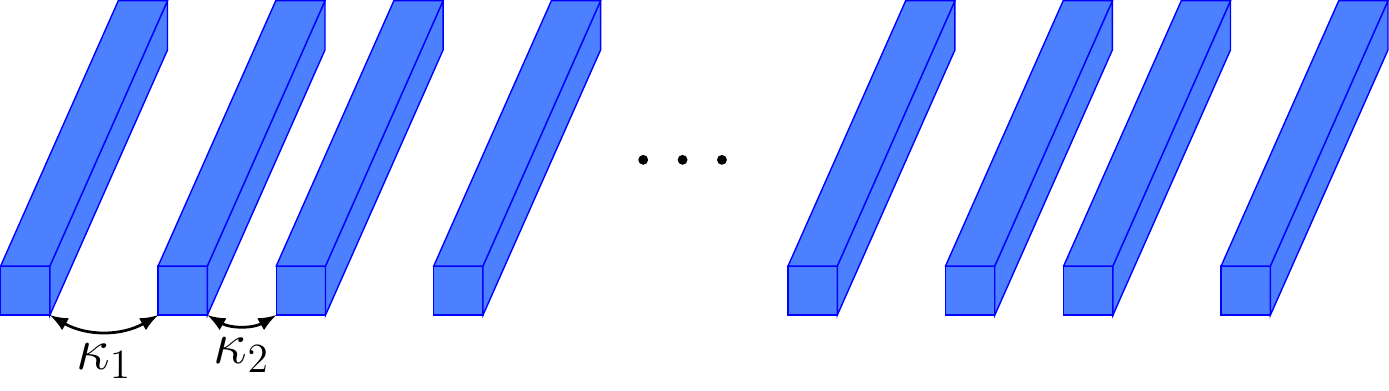}
  \caption{A diatomic SSH waveguide array with alternating coupling coefficients $\kappa_1$ and $\kappa_2$.}
  \label{fig:0}
\end{figure}

\section{SSH lattice}

Let us consider a diatomic lattice with alternating coupling coefficients, $\kappa_1$ and $\kappa_2$, as shown in Fig.~\ref{fig:0}, which is described by the Hamiltonian
\begin{multline}
  H =
  -\kappa_1\sum_{n=1}^N
  [|n,\beta\rangle\langle n,\alpha|+|n,\alpha\rangle\langle n,\beta|]
  - \\
  -\kappa_2\sum_{n=2}^N
  [|n-1,\beta\rangle\langle n,\alpha|+|n,\alpha\rangle\langle n-1,\beta|].
  \label{eq:ssh:Hamiltonian0}
\end{multline}
In the above SSH model, we account for a lattice with $N$ primitive cells and zero boundary conditions, and denote $|n,\gamma\rangle=|n\rangle\otimes|\gamma\rangle$, where $\otimes$ is the tensor product. In addition, $\gamma=\{\alpha,\beta\}$ represents the intracell internal degrees of freedom, whereas $n=1,\ldots,N$ accounts for the intercell external degrees of freedom.
We can expand the optical wave in nodal space as
\[
  |\Psi\rangle =
  \sum_{n,\gamma}C_n(z)|n,\gamma\rangle,
\]
where, depending on $\gamma$, $C=\{A,B\}$ leading to $i\dif \langle n,\gamma|\Psi\rangle/\dif z=\langle n,\gamma|H|\Psi\rangle$. Looking for stationary solutions ($C_n(z)=C_n(0)e^{-i\varepsilon z}$), we derive two coupled difference equations for the amplitudes $A_n$ and $B_n$
\begin{align}
  \varepsilon A_n & = -\kappa_2B_{n-1}-\kappa_1B_n, \label{eq:c1} \\
  \varepsilon B_n & = -\kappa_1A_n-\kappa_2A_{n+1}, \label{eq:c2}
\end{align}
where $z$ is the propagation distance and $\varepsilon$ is the eigenvalue. Combining the above equations, we can write
\begin{equation}
  (\varepsilon^2-\kappa_1^2-\kappa_2^2)C_n=
  \kappa_2\kappa_1(C_{n-1}+C_{n+1})
  \label{eq:Gamman}
\end{equation}
for the amplitudes of each sublattice. 

Even in the case of zero boundary conditions, the SSH lattice has a chiral symmetry. Specifically, if
\[
  P_\gamma = \sum_{n=1}^N|n,\gamma\rangle\langle n,\gamma|
\]
is a projection operator into sublattice $\gamma$, then we can define the chiral operator as $\Pi = P_\alpha-P_\beta$ with properties $\Pi^\dagger=\Pi$ and $\Pi^2=I$. Since $\Pi H\Pi^\dagger=-H$ or $\Pi H = -H\Pi$, for each mode $|l,+\rangle$ with positive eigenvalue $\varepsilon^{(l,+)}=\varepsilon^{(l)}>0$, a mode exists with amplitude profile $|l,-\rangle=\Pi|l,+\rangle$ and eigenvalue $\varepsilon^{(l,-)}=-\varepsilon^{(l)}$. Furthermore, as long as the eigenvalue is not zero, the power of each mode is equidistributed between the two sublattices
$\langle l,q|P_\alpha|l,q\rangle=\langle l,q|P_\beta|l,q\rangle,$
where $q=\pm$. We can expand the optical wave in modal space in terms of the supermodes 
$
  |\Psi\rangle = \sum_{l=1}^N\sum_{q=\pm}u^{(l,q)}(z) |l,q\rangle,
$
where $l$ is the mode number and $q$ determines the sign of the eigenvalue. We select to arrange the modes \begin{equation}
  |l,q\rangle=\sum_{n,\gamma}C^{(l,q)}_{n}|n,\gamma\rangle
  \label{eq:lq}
\end{equation}
according to their eigenvalues, so that $0<\varepsilon^{(1)}<\varepsilon^{(2)}<\ldots<\varepsilon^{(N)}$. When a pair of surface  modes exists, they have the lowest absolute eigenvalues and, thus, index $l=1$.

\section{Modes of the SSH lattice}

The modes of the SSH lattice were derived in~\cite{delpl-prb2011}. Here, we review the calculations for the computation of the modes and the condition for the existence of surface modes, since they are going to be useful for the theoretical and numerical calculations of the following sections. 

\subsection{Bulk modes}
Looking for sinusoidal solutions of Eq.~(\ref{eq:Gamman}) that satisfy the compatibility conditions of Eqs.~(\ref{eq:c1})-(\ref{eq:c2}) along with the boundary conditions $B_0^{(l,q)} = A_{N+1}^{(l)}=0$, we derive
\begin{align}
  A_n^{(l,q)} & = A^{(l)}(-1)^{l+1}\sin(\alpha^{(l)}(N+1-n)),\\
  B_n^{(l,q)} & = B^{(l,q)}\sin(\alpha^{(l)} n).
\end{align}
Following the calculations, we find that the eigenvalues are given by
\begin{equation}
  \varepsilon^{(l,\pm)}
  =
  \pm
  2
  \left(
    \overline\kappa^2\cos^2\frac{\alpha^{(l)}}{2}
    +
    \delta^2\sin^2\frac{\alpha^{(l)}}{2}
  \right)^{1/2},
  \label{eq:eigenvalues}
\end{equation}
where we have defined
\[
  \overline\kappa=\frac{\kappa_1+\kappa_2}{2},\quad
  \delta = \frac{\kappa_2-\kappa_1}{2},
\]
and, due to the chiral symmetry, $|A^{(l)}|=|B^{(l,q)}|$.
The coefficients $\alpha^{(l)}$ are derived from the transcendental equation
\begin{equation}
  \tan
  \left[
    \alpha^{(l)}\left(N+\frac12\right)
  \right]
  =
  \frac\delta{\overline\kappa}
  \tan\frac{\alpha^{(l)}}{2},
  \label{eq:trans}
\end{equation}
and the relative amplitudes of the two lattices are given by
\begin{equation}
  B^{(l,\pm)}=\mp A^{(l)}.
  \label{eq:relampl}
\end{equation}
The odd $\tan(\alpha^{(l)}/2)$ function on the right hand side of Eq.~(\ref{eq:trans}) is strictly increasing with limiting values $\pm\infty$ at $\pm\pi$, respectively. On the other hand, the $\tan((2N+1)\alpha^{(l)}/2)$ function of the left hand side has a period that is $2N+1$ times smaller. As a result, we expect to have at most $2N+1$ crossings (roots). However, the zero crossing is trivial (results in a zero eigenfunction). In addition, since both functions go to $\pm\infty$ at $\pm\pi$, it remains to be determined when these two crossings exist. Thus, the system supports at least $2N-2$ bulk modes. Dividing Eq.~(\ref{eq:trans}) with its left hand side, and using l'H\^opital's rule, it can be shown that the system supports $2N-2$ bulk modes, when the condition
\begin{equation}
  \frac\delta{\overline\kappa}>\frac1{1+2N}
  \label{eq:surfacemodecond}
\end{equation}
is satisfied~\cite{delpl-prb2011}. Note that Eq.~(\ref{eq:surfacemodecond}) can be written as
\[
  \frac{\kappa_2}{\kappa_1}>1+\frac1N.
\]
This latter expression is a finite $N$ refinement
of the condition $\kappa_2/\kappa_1>1$ that leads to a nontrivial Zak phase~\cite{asbot-springer2016}. When the direction of the inequality is inverted, then the lattice supports $2N$ bulk modes. 

\subsection{Surface modes}
In a similar fashion, looking for hyperbolic solutions we find
\begin{align}
  A_n^{(1,q)} & = A^{(1)}(-1)^{n+1} \sinh(\alpha^{(1)}(N+1-n)),
                \label{eq:surfacemodes01}
  \\
  B_n^{(1,q)} & = B^{(1,q)}(-1)^{n+1}\sinh(\alpha^{(1)} n), 
                \label{eq:surfacemodes02}
\end{align}
where the coefficient $\alpha^{(1)}$ is determined from the transcendental equation
\begin{equation}
    \frac{\delta}{\overline\kappa}
  \tanh\left[
    \alpha^{(1)}\left(N+\frac12\right)
  \right]
  =
  \tanh\frac{\alpha^{(1)}}2, 
  \label{eq:trans02}
\end{equation}
and, since the surface modes have the lowest absolute eigenvalues, we select $l=1$. In agreement with our previous results, dividing with the $\tanh$ term that appears on the left hand side, it can be verified that Eq.~(\ref{eq:trans02}) supports two solutions, $\alpha^{(1,\pm)} = \pm\alpha^{(1)}$ with $\alpha^{(1)}>0$, only when Eq.~(\ref{eq:surfacemodecond}) is satisfied. The respective propagation constants are given by
\begin{equation}
  \varepsilon^{(1,\pm)}=
  \pm
  (
  \kappa_1^2+\kappa_2^2-2\kappa_1\kappa_2\cosh\alpha^{(1)}
  )^{1/2},
  \label{eq:epsilonsurface}
\end{equation}
with the sublattice amplitudes satisfying Eq.~(\ref{eq:relampl}). For the rest of this paper, we choose to determine the coefficients $A^{(l)}$, $l=1,\ldots,N$, by normalizing the power of each mode to unity $\langle l,q|l,q\rangle=1$ and selecting $A^{(l)}>0$. 

In addition to the exact numerical calculation of $\alpha^{(1)}$ from Eq.~(\ref{eq:trans02}), below we derive a direct asymptotic expression, which is then utilized to determine the eigenvalues of the surface modes. We substitute Eqs.~(\ref{eq:surfacemodes01})-(\ref{eq:surfacemodes02}) to Eqs.~(\ref{eq:c1})-(\ref{eq:c2}) and set $n=1$ and $n=N$. Since $\alpha^{(1)}>0$, and approximating $\varepsilon^{(1)}\approx0$, we obtain $\kappa_2/\kappa_1\approx e^{\alpha^{(1)}}$. We look for a higher-order correction by setting
\begin{equation}
  e^{\alpha^{(1)}}=\kappa_2/\kappa_1 + u. 
  \label{eq:approx01}
\end{equation}
Substituting to Eqs.~(\ref{eq:c1})-(\ref{eq:c2}) and keeping first order terms we obtain
\begin{equation}
  u =
  -
  \left(
    \frac{\kappa_2}{\kappa_1}-
    \frac{\kappa_1}{\kappa_2}
  \right)
  \left(
    \frac{\kappa_2^N}{\kappa_1^N}-
    \frac{\kappa_1^N}{\kappa_2^N}    
  \right)^{-2}.
  \label{eq:uu}
\end{equation}
Then, from Eqs.~(\ref{eq:uu}), (\ref{eq:epsilonsurface}), we derive 
\begin{equation}
  \varepsilon^{(1)} =
  \frac{
    \kappa_2^2-\kappa_1^2
  }{
    \kappa_2
    \left(
      \frac{\kappa_2^N}{\kappa_1^N}-
      \frac{\kappa_1^N}{\kappa_2^N}
    \right)
  }
  =
  \pm
  \frac{
    \kappa_1\left(\frac{\kappa_2}{\kappa_1}-
      \frac{\kappa_1}{\kappa_2}\right)
  }{
    \frac{\kappa_2^N}{\kappa_1^N}-
    \frac{\kappa_1^N}{\kappa_2^N}    
  }.
  \label{eq:epsilon0asy}
\end{equation}
The above expression is useful in determining the propagation constant (and thus, as we will see, the coupling period) directly, as a function of the coupling coefficients and the length of the lattice. It is found to be in very good agreement with the numerical solution of the transcendental Eq.~(\ref{eq:trans02}). Note that the surface supermodes of the SSH lattice are known as ``near-zero modes'', because their eigenvalues are close to zero.

\section{Coupled mode theory}

For the rest of this paper, we assume that the condition given by Eq.~(\ref{eq:surfacemodecond}) is satisfied, and thus the lattice supports two surface modes.
We are going to develop a coupled-mode theory formalism that incorporates all the modes of the SSH lattice (both bulk and surface). We select to utilize a local mode decomposition of the form
\begin{align}
  |l,\alpha\rangle = & \frac{|l,+\rangle+|l,-\rangle}{\sqrt{2}},
                       \label{eq:localbasis1} \\ 
  |l,\beta\rangle = & \frac{(-1)^{N+l+1}( |l,+\rangle-|l,-\rangle)}{\sqrt{2}}.
                      \label{eq:localbasis2}
\end{align}
In Eqs.~(\ref{eq:localbasis1})-(\ref{eq:localbasis2}), the local mode $|l,\gamma\rangle$ excites only the $\gamma$ sublattice, and the coefficient $1/\sqrt{2}$ is used for normalization purposes. Note that, even in this decomposition, the modes are orthogonal $\langle l',\gamma'|l,\gamma\rangle=\delta_{l,l'}\delta_{\gamma,\gamma'}$.
In the basis given by Eqs.~(\ref{eq:localbasis1})-(\ref{eq:localbasis2}), the surface modes or zero modes $|l=1,\gamma\rangle$ are exponentially localized, one on the left ($\gamma=\alpha$) and one on the right ($\gamma=\beta$) side of the lattice. We expand the optical wave as
\[
  |\Psi\rangle = \sum_{l=1}^N\sum_{\gamma=\{\alpha,\beta\}}c^{(l)}(z)|l,\gamma\rangle,
\]
where $c=\{a,b\}$, depending on the sublattice, and derive the following coupled-mode theory equations for the wave amplitudes
\begin{align}
  i\frac{\dif}{\dif z}a^{(l)}(z) & =-\varepsilon^{(l)}b^{(l)},
  \label{eq:cmt1} \\
  i\frac{\dif}{\dif z}b^{(l)}(z) & =-\varepsilon^{(l)}a^{(l)}.
  \label{eq:cmt2}
\end{align}
The solution of Eqs.~(\ref{eq:cmt1})-(\ref{eq:cmt2}) leads to periodic sinusoidal oscillations between $|l,\alpha\rangle$ and $|l,\beta\rangle$, with period equal to
\begin{equation}
  L^{(l)} = \frac\pi{\varepsilon^{(l)}}.
  \label{eq:period}
\end{equation}
Note that we can define the coupling lengths as half of the respective periods, $L^{(l)}/2$.

\section{Edge excitation}

From Eqs.~(\ref{eq:cmt1})-(\ref{eq:cmt2}) with $l=1$, we see that when the zero edge modes are excited the power oscillates periodically between them without leakage into the bulk modes. However, experimentally, it is not very simple to engineer the initial condition to selectively excite only the surface modes. Thus, it is important to ask how much of the power is going to go to the surface modes, when simpler excitation schemes are used. In particular, we consider perhaps the simplest case, where only a single waveguide on the left edge of the lattice is excited on the input plane
\begin{equation}
  |\Psi(0)\rangle =
  |n=1,\alpha\rangle.
  \label{eq:singlewaveguide}
\end{equation}
It is straightforward to see that the amount of power that excites the surface mode $|l=1,\alpha\rangle$ is
\begin{equation}
  P_{1,\alpha} = 
  |\langle l=1,\alpha|\Psi(0)\rangle|^2 = 2(A^{(1)})^2\sinh^2(a^{(1)}(N+1)), 
  \label{eq:P1alpha}
\end{equation}
whereas the power of the bulk mode $|l>1,\alpha\rangle$ is given by
\begin{equation}
  P_{l,\alpha} = 
  |\langle l>1,\alpha|\Psi(0)\rangle|^2 = 2(A^{(l)})^2\sin^2(a^{(1)}(N+1)).
  \label{eq:Plalpha}
\end{equation}
On the other hand, the modes of sublattice $\beta$ are not excited on the input plane $P_{l,\beta}=0$ ($\langle l,\beta|\Psi(0)\rangle =0$).

\begin{figure}
  \centering
  \includegraphics[width=\columnwidth]{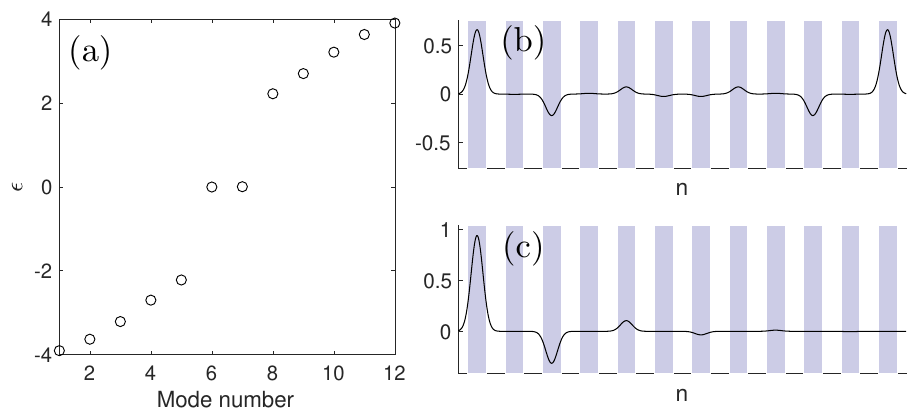}
  \caption{(a) Eigenvalues of an SSH lattice with coupling coefficients $\kappa_1=1$ and $\kappa_2=3$, consisting of $N=6$ primitive cells.
    (b) The amplitude profile of the near-zero surface mode with a positive eigenvalue $|l=1,+\rangle$.
    (c) The amplitude profile of the left zero surface mode that excites only the $\alpha$ sublattice $|l=1,\alpha\rangle$.
  }
  \label{fig:1}
\end{figure}

\section{Linear dynamics}
Below, we present numerical results of an SSH lattice with $N=6$ primitive cells and coupling coefficients $\kappa_1=1$ and $\kappa_2=3$. Using these parameters, we obtain $\varepsilon^{(1)}\approx0.00366$ and thus $L^{(1)}\approx859$. Reducing the coupling contrast $\kappa_2/\kappa_1$ or the size of the lattice $N$, leads to the reduction of the period of the oscillations. For example, for $N=5$ and the same coupling coefficients, we obtain $L^{(1)}\approx286$. In Fig.~\ref{fig:1}(a), we depict the eigenvalues of the SSH lattice with zero boundary conditions, in increasing eigenvalue order. The bulk eigenvalues form two bands, whereas the surface modes are located close to the middle of the gap. In addition, in Fig.~\ref{fig:1}(b), we can see the amplitude profile of the near zero surface mode with positive eigenvalue $|l=1,+\rangle$. The second supermode $|l=1,-\rangle$ is derived by application of the chiral operator, leading to opposite amplitudes for the $\beta$ sublattice. Thus, since $|l=1,+\rangle$ is even, $|l=1,-\rangle$ is odd. The left zero mode that excites the $\alpha$ sublattice, $|l,\alpha\rangle$, is shown in Fig.~\ref{fig:1}(c).

\begin{figure}
  \centering
  \includegraphics[width=\columnwidth]{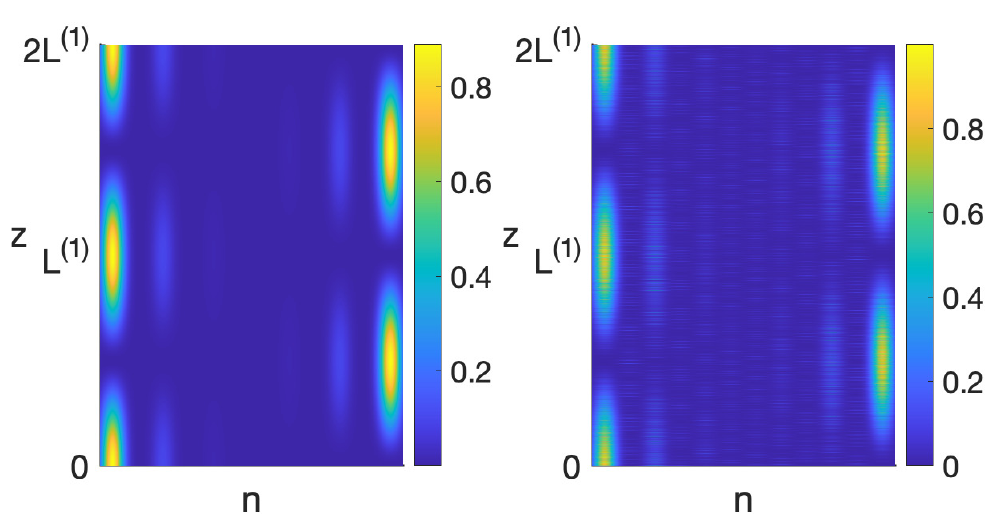}
  \caption{Intensity dynamics depicting the periodic oscillations between the two zero surface modes. In (a) the left zero surface mode that excites the $\alpha$ sublattice, $|l=1,\alpha\rangle$, is launched on the input plane, while in (b) the first waveguide that resides on the left edge of the array, $|n=1,\alpha\rangle$, is excited. Only minor differences are observed between the two simulations. The parameters are the same as in Fig.~\ref{fig:1}.}
  \label{fig:2}
\end{figure}
We numerically integrate the SSH lattice by utilizing a Runge-Kutta scheme. In the simulation shown in Fig.~\ref{fig:2}(a), we launch the left zero mode, $|l=1,\alpha\rangle$, which is localized on the left edge of the array. The theoretical value for the period of the oscillations between the two surface modes, is verified by the numerical results. In Fig.~\ref{fig:2}(b) we use the, experimentally simpler, excitation scheme, where only the waveguide on the left edge of the lattice, $|n=1,\alpha\rangle$, is initially exited. Interestingly, we only see minor differences when comparing these two simulations. In particular, in Fig.~\ref{fig:2}(b), we notice some low intensity and high frequency noise in between the two surface modes.

\begin{figure}
  \centering
  \includegraphics[width=0.55\columnwidth]{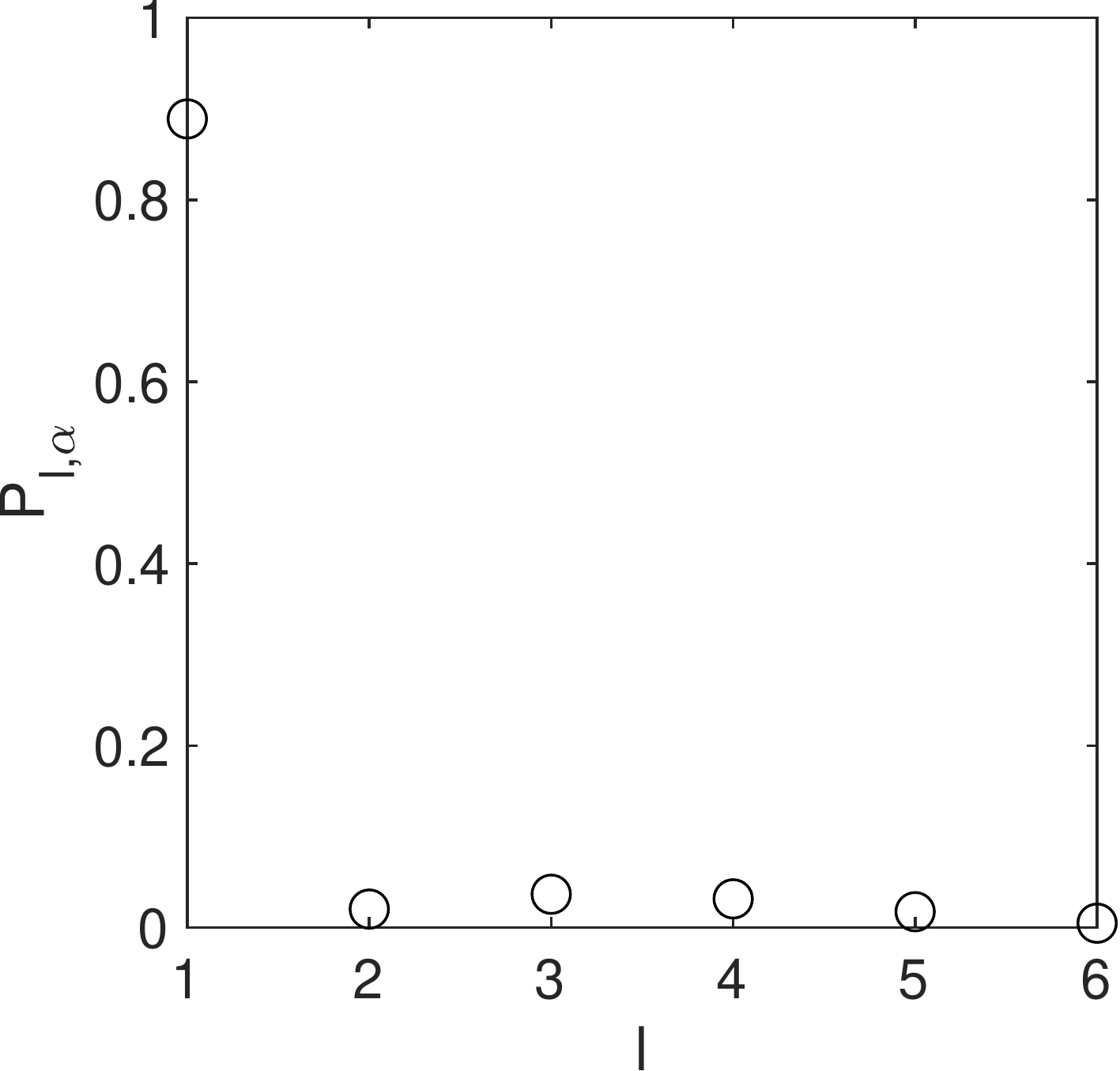}
  \caption{Initial distribution of power $P_{l,\alpha}$ between the modes of the SSH lattice $|l,\alpha\rangle$, when only the first (left) waveguide is excited. The lattice parameters are the same as in the previous figures.}
  \label{fig:3}
\end{figure}
To quantify the comparison between the two simulations shown in Fig.~\ref{fig:2}, we decompose the initial profile, in the case where only the left waveguide of the lattice is excited, into the sublattice modes $|l,\alpha\rangle$. As we have shown, the modes of sublattice $\beta$ ($|l,\beta\rangle$) are not involved ($P_{l,\beta}=0$).
In Fig.~\ref{fig:3}, we see the distribution of power between the modes of the $\alpha$ sublattice, $P_{l,\alpha}$, which is given by Eqs.~(\ref{eq:P1alpha})-(\ref{eq:Plalpha}). Importantly, about $89\%$ of the power excites the surface mode $|l=1,\alpha\rangle$. The rest of the power is distributed between the bulk modes, with the highest amount, about $3.7\%$, exciting the mode $|l=3|\alpha\rangle$. Then, during propagation, the power oscillates in pairs, between the $|l,\alpha\rangle$ and the $|l,\beta\rangle$ modes, with each oscillation having its own spatial frequency. Since $\varepsilon^{(l>1)}\gg\varepsilon^{(1)}$, the low-intensity noise generated by the excitation of the bulk modes is high frequency, as compared to the oscillations between the surface modes. 

\section{Nonlinear dynamics}

Assuming that the waveguide lattice is Kerr nonlinear then, in nodal space, the dynamics of the SSH lattice satisfy
\begin{align}
  i\frac{\dif a_n}{\dif z}
  +\kappa_2b_{n-1}+\kappa_1b_{n}+g |a_n|^2a_n & =0, \label{eq:nl11}
  \\
  i\frac{\dif b_n}{\dif z}
  +\kappa_1a_{n}+\kappa_2a_{n+1}+g |b_n|^2b_n & =0. \label{eq:nl12}
\end{align}
On the other hand, we can follow a modal approach similar to the one used in the linear case. This can be done by expanding the optical wave into the sublattice modes, where the effect of the Kerr nonlinearity results in additional wave mixing terms. The resulting modal system of equations reads
\begin{equation}
  i\frac{\dif c^{(l)}}{\dif z}+\varepsilon^{(l)}\tilde c^{(l)}
  +\gamma\sum_{l_1,l_2,l_3}
  Q_{l,l_1,l_2,l_3}(c^{(l_1)})^*c^{(l_2)}c^{(l_3)}=0,
  \label{eq:modal}
\end{equation}
where $c^{(l)}=\{a^{(l)},b^{(l)}\}$ is the $z$-dependent amplitude of mode $l$ that excites the sublattice $c$, $\tilde c=b$ when $c=a$ and vice versa, 
\[
  Q_{l,l_1,l_2,l_3} =
  \sum_{n}
  (C^{(l_1)}_n)^*(C^{(l)}_n)^*C^{(l_2)}_nC^{(l_3)}_n,
\]
and $C^{(l)}_n=\{A^{(l)}_n,B^{(l)}_n\}$ are the field profiles of modes $|l,\alpha\rangle$ and $|l,\beta\rangle$, respectively. Note that each sum in Eqs.~(\ref{eq:modal}) involve $N^3$ terms. Thus, the modal system of equations is numerically far more complicated to solve in comparison to the original nodal system. Focusing on the case of surface mode excitation, then a significant simplification of Eqs.~(\ref{eq:modal}) arises, by ignoring the nonlinear contributions that do not play a significant role in the dynamics. These are the small wave mixing terms and/or the asynchronous terms between the surface ($l=1$) and the bulk modes or the surface modes ($l\ge1$). Thus, we obtain the following coupled equations for the surface modes
\begin{align}
  i\frac{\dif a^{(1)}}{\dif z}+\kappa b^{(1)}+G |a^{(1)}|^2a^{(1)} & =0, \label{eq:nl21} \\
  i\frac{\dif b^{(1)}}{\dif z}+\kappa a^{(1)}+G |b^{(1)}|^2b^{(1)} & =0, \label{eq:nl22}
\end{align}
that generalize the linear system described by Eqs.~(\ref{eq:cmt1})-(\ref{eq:cmt2}). In Eqs.~(\ref{eq:nl21})-(\ref{eq:nl22}) $\kappa=\varepsilon^{(1)}$ is the coupling coefficient, the effective nonlinearity is given by $G=gQ$, 
\begin{equation}
  Q = \sum_n(A_n^{(1)})^4=\sum_n(B_n^{(1)})^4
  \label{eq:G}
\end{equation}
and $A_n^{(1)}$, $B_n^{(1)}$ are the field profiles of the zero (sublattice) surface modes. Note that Eqs.~(\ref{eq:nl21})-(\ref{eq:nl22}) can be solved analytically~\cite{jense-jqe1982}. Equations~(\ref{eq:modal}) are far more complicated than the reduced system, which is derived by ignoring a variety of nonlinear terms. Thus, it is important to examine the accuracy and regimes of validity of Eqs.~(\ref{eq:nl21})-(\ref{eq:nl22}).

\begin{figure}
  \centering
  \includegraphics[width=\columnwidth]{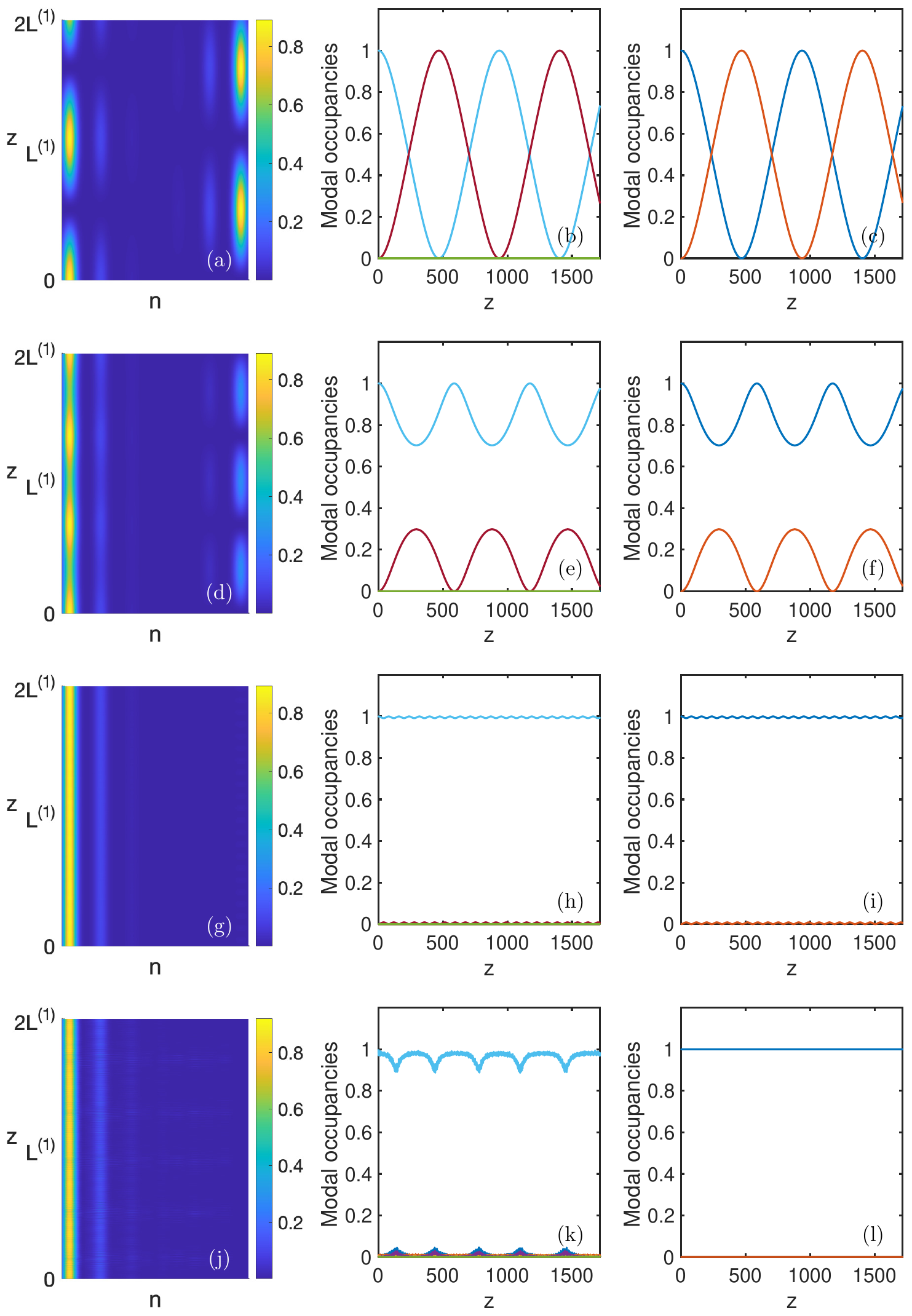}
  \caption{Dynamics of an SSH lattice when the Kerr nonlinear coefficient is $g=0.01$, $0.02$, $0.1$, and $1$ in the first, second, third, and fourth row, respectively.  In the fist column the intensity dynamics as a function of the propagation distance is shown when the left zero surface mode is initially excited. In the second column the modal power occupancies vs the propagation distance $z$, derived by expanding the amplitude profile into a basis that consists of the sublattice modes $|l,\gamma\rangle$, is presented. In the last column, the surface mode dynamics is depicted by utilizing the simplified coupled-mode theory Eqs.~(\ref{eq:nl21})-(\ref{eq:nl22}) that does not account for the surface modes. Note that the comparison between the second and the third column is excellent, except in the last row where, due to the relatively strong nonlinearity, bulk modes are excited.}
  \label{fig:4}
\end{figure}
In the remaining section, we perform a series of numerical simulations, to examine the dynamics of Eqs.~(\ref{eq:nl11})-(\ref{eq:nl12}) [which are equivalent to  Eqs.~(\ref{eq:modal})], when the left surface mode is initially excited. In addition, we are going to compare these results with the solution of the simplified coupler model~(\ref{eq:nl21})-(\ref{eq:nl22}). The lattice parameters are kept the same as in the linear case ($N=6$, $\kappa_1=1$, $\kappa_2=3$), which leads to $Q=0.8$. In all our simulations on the SSH lattice, we have included additional random noise, to trigger four-wave mixing terms that can lead to the excitation of the bulk modes.

In Fig.~\ref{fig:4}(a)-(c), the nonlinear coefficient is small $g=0.01$. We clearly see in Fig.~\ref{fig:4}(a) the oscillations between the two surface modes. The nonlinearity slightly modifies the period and the amplitude of the oscillations, in comparison to the linear limit. Furthermore, we are not able to detect any noticeable excitation of bulk modes due to the presence of nonlinearity. This can be more clearly demonstrated by decomposing the amplitude profile into the bulk and the surface sublattice $|l,\gamma\rangle$ modes of the lattice. In particular, in Fig.~\ref{fig:4}(b), we can see that the amplitudes of the two zero surface modes exhibit periodic oscillations. The power exchange between the two surface modes is almost 100\%, whereas the amplitude of the bulk modes is almost zero. In Fig.~\ref{fig:4}(c) we depict the dynamics of the two surface zero modes, by solving the simplified coupled-mode theory Eqs.~(\ref{eq:nl21})-(\ref{eq:nl22}). We note that the comparison with Fig.~(\ref{fig:4}(b) is excellent.

\begin{figure}
\centerline{
\includegraphics[width=0.9\columnwidth]{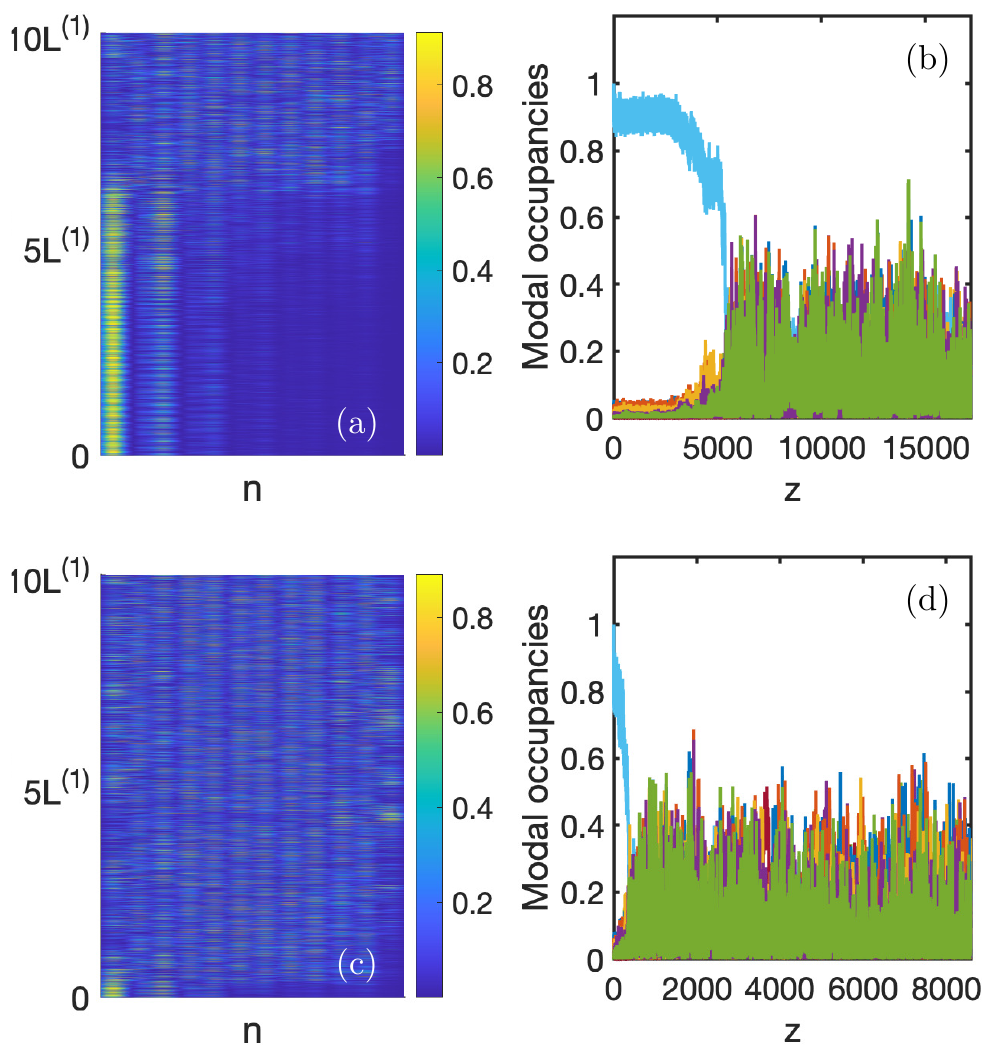}
}
\caption{Dynamics of an SSH lattice in the case of strong Kerr nonlinearity $g=2$ and $3$ in the first and second row, respectively. In the fist column the intensity dynamics as a function of the propagation distance is shown when the left zero surface mode is initially excited. In the second column the modal power occupancies vs the propagation distance $z$, derived by expanding the amplitude profile into a basis that consists of the sublattice modes $|l,\gamma\rangle$, are presented.}
\label{fig:5}
\end{figure}

In Fig.~\ref{fig:4}(d)-(f) the nonlinear coefficient is increased to $\gamma=0.02$. We can see periodic oscillations between the left and the right surface modes. However, in this case only a small part of the total power is transferred from the left to the right surface mode. We also observe that the excitation of the bulk modes is still very small and the comparison with the simplified coupler model is excellent. Once we increase the nonlinearity to $\gamma=0.1$ [Fig.~\ref{fig:4}(g)-(i)] almost all the power remains trapped into the left zero surface mode. In particular, the bulk modes are effectively not excited, and we observe some minute high frequency oscillations between the two zero surface modes. Interestingly, even in this case, the coupler model is in very good agreement in predicting the period and small amplitude oscillations between the two surface modes.

In Fig.~\ref{fig:4}(j)-(l) the nonlinearity is increased to $\gamma=1$. During propagation most of the power remains in the left surface mode. However, the main difference in comparison to the previous simulations, is that the presence of relatively strong nonlinearity induces the excitation of the bulk modes of the system. Specifically, in Fig.~\ref{fig:4}(k) we observe that, quasi-periodically, a small portion of the total power of the left surface zero mode is transferred to the bulk modes. Of course, this behavior can not be captured by the simplified coupler equations in Fig.~\ref{fig:4}(l) that do not account for the bulk modes. However, since the portion of the total power that is transferred to the bulk modes is small, we can still claim that the reduced model can be utilized to provide a relatively accurate estimate of the wave dynamics.

If we further increase the nonlinearity, as we can see in Fig.~\ref{fig:5}, then strong interactions between the bulk and the surface modes take place. The power is gradually transferred to all the modes of the system leading to a chaotic behavior. For example, in the first row of Fig.~\ref{fig:5} where $\gamma=2$, we see that, in the initial stages, most of the power remains in the left surface mode. However, gradually the power is transferred to all the modes of the system. Eventually, at $z\approx 7L^{(1)}$ there is no signature of the highlighted presence of the surface modes. In general, as the nonlinearity is increased the transition to chaotic behavior is going to take place faster. For example, in the second row of Fig.~\ref{fig:5} where $\gamma=3$ the transition takes place at $z\approx L^{(1)}$. The dynamics of the SSH lattice after this point might be examined using, for example, a thermodynamic approach.

\section{Conclusions}

In conclusion, we have developed a coupled-mode theory approach for the sublattice modes of an SSH lattice. The theory can be applied to analytically describe the coupling behavior between the two zero surface modes of the system. We have found that launching light on the waveguide that is located on the left (or the right) edge of the array, can be very efficient in exciting the surface modes. Such an excitation scheme is very simple to experimentally implement. We extended our analysis in the nonlinear regime by including the effect of cubic (Kerr) nonlinearity. Our simplified model is able to capture the dynamics when the nonlinearity is small or moderate. On the other hand, strong nonlinearity results to the quasi periodic excitation of the bulk modes and eventually to chaotic behavior with no actual signature of the presence of surface modes. We expect that our results can be generalized to different types of lattices, including systems with broken sublattice symmetry provided that nontrivial edge states are supported.
\\\\

\newcommand{\noopsort[1]}{} \newcommand{\singleletter}[1]{#1}

\end{document}